\documentclass[aps,twocolumn,final,superscriptaddress,
nofootinbib,nopacs,
nokeys,twoside,notitlepage]{revtex4-2}

\usepackage[utf8]{inputenc}
\usepackage{lmodern}

\usepackage{amsmath}
\usepackage{amssymb}
\usepackage{mathtools}

\def\square{\kern1pt\vbox{\hrule height 0.4pt\hbox{\vrule width 0.4pt\hskip 3pt
   \vbox{\vskip 6pt}\hskip 3.pt\vrule width 0.9pt}\hrule height 0.9pt}\kern1pt}

\usepackage{tensor}

\newcommand{\scr}{\scriptscriptstyle}

\newcommand{\longsim}{\scalebox{1.8}[1]{$\sim$}}

\usepackage{xcolor}

\usepackage{upgreek}

\usepackage[normalem]{ulem}

\usepackage{cancel}

\usepackage[hidelinks]{hyperref}

\begin{document}
\title{Even the photon propagator must break de Sitter symmetry}

\author{Dra\v{z}en Glavan}
\email[]{glavan@fzu.cz}
\affiliation{CEICO, Institute of Physics of the Czech Academy of Sciences,
	Na Slovance 1999/2, 182 21 Prague 8, Czech Republic}
\author{Tomislav Prokopec}
\email[]{t.prokopec@uu.nl}
\affiliation{Institute for Theoretical Physics, Spinoza Institute \& EMME$\Phi$,
	Utrecht University, Buys Ballot Building, Princetonplein 5,
	3584 CC Utrecht, The Netherlands}

\begin{abstract}
The propagator for the massless vector field in de Sitter space cannot maintain de Sitter invariance 
in the general covaraint gauge, except in the exactly transverse gauge limit. This is due to a previously overlooked
Ward-Takahashi identity that the propagator must satisfy. Here we
construct the 
propagator that satisfies all the conditions of a consistently quantized theory. Our solution
preserves cosmological 
symmetries and dilations, but breaks spatial special conformal transformations. The solution amounts
to adding a homogeneous de Sitter breaking term to previously reported de Sitter invariant solutions
of the propagator equation
of motion. Even though the corrections we report pertain to the 
gauge sector of the linear theory, 
they are relevant and have to be accounted for when interactions are included.
\end{abstract}

\maketitle

\noindent{\bf Introduction.} 
Understanding interacting quantum field theory in de Sitter space is of paramount importance 
for apprehending the physics of the primordial inflationary phase of the Universe.
Computations of quantum loop corrections in realistic slow-roll inflation are still prohibitively
difficult, so simplifications are necessary. The de Sitter space is often taken as an appropriate 
idealization for two reasons: (i) it is close enough to slow-roll inflation that is 
phenomenologically relevant, and (ii) it is a maximally symmetric space.
It is the latter that is very often useful when describing physical systems -- 
more symmetric they are the simpler the description. Such is our experience
in Minkowski space, where Poincar\'e invariance 
provides an efficient organizational principle for computations.
It is often assumed that de Sitter symmetries provide the same
level of simplifications and organize the computations in an
economical manner. Even though adhering to symmetries is the right approach
in many circumstances, it must not be taken for granted in de 
Sitter space.

Two-point functions of free fields are essential ingredients for perturbative computations in 
quantum field theory. In maximally symmetric spaces it is natural to assume
that they respect symmetries of the background spacetime.
However, it has long been known that
issues with this approach arise already for arguably the simplest
system of minimally coupled, massless scalar (MMCS),
whose propagator satisfies the equation of motion,
%
\begin{equation}
\sqrt{-g} \, \square \, i \Delta(x;x') = i \delta^D(x\!-\!x') \, ,
\end{equation}
where $\square \!=\! g^{\mu\nu}\nabla_{\!\mu} \nabla_{\!\nu}$ denotes the d'Alembertian (wave) 
operator. Even though this equation is invariant under de Sitter symmetries, it does not
admit a de Sitter invariant solution with the appropriate singularity structure~\cite{Allen:1985ux,Allen:1987tz}
in any number~$D$ of spacetime dimensions.
This letter is devoted to pointing out a similar, but more subtle, obstruction to 
maintaining de Sitter symmetry for the massless vector propagator in 
the general covariant gauge.
Even though the equation of motion does allow 
a de Sitter invariant solution,
it is the Ward-Takahashi identity, overlooked thus far
for propagators of massless vector fields, 
that prevents a de Sitter invariant solution. Here we
present a solution for the photon 
propagator in~$D$-dimensional spacetime, appropriate for dimensionally regulated 
quantum loop computations, that accounts for both the 
equation of motion and the 
Ward-Takahashi identity.

\bigskip
\noindent{\bf Photon in the general covariant gauge.} 
The physical photon is conformally coupled to gravity in four dimensions,
which is no longer true in~$D$-dimensional spacetime,
\begin{equation}
S[A_\mu] = \int\! d^{D\!}x \, \sqrt{-g} \, 
	\biggl[
	- \frac{1}{4} g^{\mu \rho} g^{\nu \sigma} F_{\mu\nu} F_{\rho\sigma}
	\biggr]
	\, ,
\label{action}
\end{equation}
where~$F_{\mu\nu} \!=\! \partial_\mu A_\nu \!-\! \partial_\nu A_\mu$ is the 
field strength tensor of the vector potential~$A_\mu$.
We consider the spacetime to be the expanding Poincar\'e patch of de Sitter 
where the metric,~$g_{\mu\nu}\!=\!a^2(\eta)\eta_{\mu\nu}$,
is conformal to the Minkowski space metric,~$\eta_{\mu\nu}\!=\!{\rm diag}(-1,1,\dots,1)$,
$g \!=\! {\rm det}(g_{\mu\nu})$, and
$a(\eta) \!=\! 1/[1 \!-\! H(\eta \!-\! \eta_0)]$ is the scale factor 
expressed in terms of conformal time $\eta$ and a constant Hubble parameter $H$.
We shall not consider symmetry breaking theories, in which the vector field can acquire a mass,
nor shall we consider spatially compact global coordinates on de Sitter 
where the problem of linearization
instability arises~\cite{Miao:2009hb}.

Quantization of the theory requires fixing a gauge. 
Choosing the general covariant gauge,
\begin{equation}
S_{\rm gf}[A_\mu] = \int\! d^{D\!}x \, \sqrt{-g} \, 
	\biggl[
	- \frac{1}{2\xi} \bigl( \nabla^\mu \! A_\mu \bigr)^{\!2}
	\biggr]
	\, ,
\label{action: gf}
\end{equation}
allows to maintain de Sitter symmetries of the dynamics. However, it breaks conformal
coupling of the photon even in four spacetime dimensions, and for any choice of 
the gauge-fixing parameter~$\xi$.

Central objects in perturbative nonequilibrium quantum field theory 
are the two-point functions determined from the free theory.
The relevant ones are the Feynman propagator and the positive frequency 
Wightman function,
\begin{align}
i \bigl[ \tensor*[_\mu]{\Delta}{_\nu} \bigr](x;x')
	={}&
	\bigl\langle \Omega \bigr| \mathcal{T} \big[\hat{A}_\mu(x) \hat{A}_\nu(x')\big] \bigl| \Omega \bigr\rangle
	\, ,
\label{propagator: definition}
\\
i \bigl[ \tensor*[_\mu^{\scr - \! }]{\Delta}{_\nu^{\scr \!+}} \bigr](x;x')
	={}&
	\bigl\langle \Omega \bigr| \hat{A}_\mu(x) \hat{A}_\nu(x') \bigl| \Omega \bigr\rangle
	\, ,
\label{positive frequency Wightman function}
\end{align}
where $\bigl| \Omega \bigr\rangle$ is the state,  $\mathcal{T}$ stands for time ordering, and
$\hat{A}_\mu(x)$ is the free photon field operator.
The propagator equation of motion in this gauge is,
\begin{equation}
\sqrt{-g} \, \mathcal{D}^{\mu\nu}
	\, i \bigl[ \tensor*[_\nu]{\Delta}{_\alpha} \bigr](x;x')
	=
	\delta^\mu_{\alpha} \, i \delta^{D}(x\!-\!x')
	\, ,
\label{EOMs}
\end{equation}
where we set $\hbar =1$,
and the kinetic operator is,
\begin{equation}
\mathcal{D}^{\mu\nu} = g^{\mu\nu} \square - \Bigl( 1 \!-\! \frac{1}{\xi} \Bigr) \nabla^\mu \nabla^\nu
	- R^{\mu\nu}
	\, ,
\label{photon operator}
\end{equation}
where $R^{\mu\nu} \!=\! (D\!-\!1)H^2 g^{\mu\nu}$ is the Ricci tensor in de Sitter.
Due to the exchange symmetry $(\mu,x)\leftrightarrow (\nu,x')$,
the photon propagator~(\ref{propagator: definition}) obeys an equation analogous 
to~(\ref{EOMs}) on the other leg $(\nu, x')$.
The Wightman function obeys the same equation of motion~(\ref{EOMs}), 
but without the local term on the right-hand-side.

The photon two-point functions
 in covariant gauges~(\ref{action: gf})
have been considered in several works over the last decades, starting from the
seminal work of Allen and Jacobson~\cite{Allen:1985wd}, who reported many results, among which
the de Sitter space covariant gauge propagator for~$\xi\!=\!1$ in~$D$ spacetime dimensions.
Subsequent works have extended and generalized this result. Tsamis and Woodard~\cite{Tsamis:2006gj}
reported the transverse massive vector propagator, whose massless limit 
reduces to the photon propagator
in the Landau gauge ($\xi\!\to\!0$) in~$D$ dimensions; 
Youssef~\cite{Youssef:2010dw} 
computed the propagator for arbitrary~$\xi$ in~$D\!=\!4$ spacetime dimensions; 
and Fr\"ob and Higuchi~\cite{Frob:2013qsa} reported the result for 
a massive vector propagator, with the massless limit producing the 
photon propagator for
arbitrary~$\xi$ and arbitrary~$D$.
The last of these encompasses all the previously reported results as special cases.
Only propagators from~\cite{Allen:1985wd} and~\cite{Tsamis:2006gj} were used for loop
computations in de Sitter. The former was rederived and used to study scalar 
electrodynamics~\cite{Kahya:2006ui}, while the latter was utilized for both
scalar electrodynamics~\cite{Prokopec:2006ue,Prokopec:2008gw,Kahya:2006ui,Prokopec:2007ak}, 
and for quantum 
gravity interacting with electromagnetism~\cite{Glavan:2015ura,Glavan:2016bvp}.

In this letter we are interested only in the massless vector (photon) propagators.
All of the photon propagators reported in previous works satisfy the equation of motion~(\ref{EOMs}).
However, it was pointed out recently in~\cite{Glavan2022} that photon propagators should satisfy 
additional subsidiary conditions dictated by the consistent canonical quantization in
average/multiplier gauges. Among them is the condition that the double divergence of
both the Feynman propagator and the Wightman function should vanish 
off-coincidence. But the reported results violate this 
condition,~\footnote{ The Wightman function exhibits the same problem as in~(\ref{problem}), 
except that the local term on the right-hand-side is absent.
}
\begin{equation}
\nabla^\mu \nabla'^\nu i \bigl[ \tensor*[_\mu]{\Delta}{_\nu} \bigr](x;x')
	\overset{!}{=}- \xi \frac{ i \delta^D(x \!-\! x') }{ \sqrt{-g} }
	- \xi\frac{ H^{D} \, \Gamma(D) }{ (4\pi)^{\frac{D}{2} } \, \Gamma\bigl( \frac{D}{2} \bigr) } \, ,
\label{problem}
\end{equation}
except in the limit~$\xi\!\to\!0$ when the offending term vanishes, suggesting than only the 
Tsamis-Woodard result is consistent.
This is a problem that needs to be addressed.
In the companion paper~\cite{Glavan2022_2} to this letter we consider the problem
from the
first principles of canonical quantization and construct the photon propagator
as a sum over modes.
This letter is devoted to resolving the problem in~(\ref{problem})
in an elegant manner by considering the 
Becchi-Rouet-Stora-Tyutin (BRST) quantization~\cite{Kugo:1979gm}.
Our approach is similar to the one employed in~\cite{Frob:2017gez}
for studying retarded and advanced Green's functions for massive vector fields.
However, the problem in~(\ref{problem}) is not encountered for these Green's functions,
as for them the problematic term is absent.

\bigskip

\noindent {\bf BRST quantization.} 
In addition to introducig the gauge-fixing term~(\ref{action: gf}),
the BRST formalism
requires the inclusion of the Faddeev-Popov ghost action for Grassmann 
fields~$c$ and~$\overline{c}$,
\begin{equation}
S_{\rm gh}[c, \overline{c}]
	= \int\! d^{D\!}x \sqrt{-g}\, 
		g^{\mu\nu}\bigl( \nabla_\mu \overline{c} \bigr)
		\bigl( \nabla_\nu c \bigr)
		\, ,
\label{Grassmann action}
\end{equation}
to the complete gauge-fixed action,~$S_\star \!=\! S \!+\! S_{\rm gf} \!+\! S_{\rm gh}$, 
which is invariant under infinitesimal BRST transformations,
\begin{equation}
A_\mu \to A_\mu + \theta \xi \partial_\mu c \, ,
\quad \
\overline{c} \to \overline{c}  -  \theta \nabla^\mu A_\mu\,,
\quad \
c  \to c
\, ,
\end{equation}
parametrized by~$\theta$, that are
generated by the associated conserved BRST charge,
\begin{equation}
{\rm Q} = \int\! d^{D-1}x \, a^{D-4} 
	\Bigl[ 
		a^2 \bigl( \nabla^\mu \! A_\mu \bigr) \partial_0 c
		+ \xi F_{0i} \partial_i c \, 
		\Bigr]
	\, .
\label{ghost charge Q}
\end{equation}
The ghost 
propagator, $i\Delta_{c}(x;x') \!=\! \bigl\langle\Omega\bigl| \mathcal{T} \, \bigl[\hat{\overline{c}}(x) \hat{c}(x')\bigr] 
\bigr|\Omega\bigr\rangle$,
satisfies,
\begin{equation}
\sqrt{-g} \, \square \, i \Delta_{c}(x;x') = 
	i \delta^D(x\!-\!x')
\, .
\label{ghost EOMs}
\end{equation}
This equation implies that the ghost propagator equals the MMCS 
propagator,~$i \Delta_{c}(x;x') \!=\! i \Delta(x;x')$.
Thus, the ghost propagator
must break de Sitter symmetry. 
The natural choice for the MMCS propagator in
the Poincar\'{e} patch of de Sitter is the one preserving cosmological symmetries,
but breaking dilations and special spatial conformal transformations~\cite{Onemli:2002hr}.
For our purposes it is best to write it as a limit~\cite{Janssen:2008px},
\begin{equation}
i\Delta(x;x')
                = \lim_{\lambda\to\nu+1} i \Delta_{\lambda}(x;x')
	\, ,
	\qquad
	\nu = \frac{D\!-\!3}{2}
	\, ,
\label{MMCS propagator}
\end{equation}
of a propagator with an effectively slightly tachyonic 
mass~$M^2 \!=\! [(D\!-\!1)^2/4 \!-\! \lambda^2] H^2 \!<\! 0$,
\begin{equation}
i \Delta_{\lambda}(x;x')
	=
	\mathcal{F}_{\lambda}(y) 
	+ \mathcal{W}_{\lambda}(u)
	\, ,
\label{dS breaking 1}
\end{equation}
which consists of a de Sitter invariant part,
\begin{align}
\mathcal{F}_\lambda(y)
	={}& \frac{H^{D-2}}{ (4\pi)^{\frac{D}{2}} } 
	\frac{ \Gamma\bigl( \frac{D-1}{2} \!+\! \lambda \bigr) \, \Gamma\bigl( \frac{D-1}{2} \!-\! \lambda \bigr) }{ \Gamma\bigl( \frac{D}{2} \bigr) }
\nonumber \\
&	\hspace{0.5cm}
	\times
	{}_2F_1\Bigl( \tfrac{D-1}{2} \!+\! \lambda , \tfrac{D-1}{2} \!-\! \lambda, \tfrac{D}{2} , 1 \!-\! \tfrac{y}{4} \Bigr) \, ,
\label{F def}
\end{align}
dependent on a de Sitter invariant 
distance,
\begin{equation}
y \!=\! aa' H^2 \Bigl[ \| \vec{x} \!-\! \vec{x}^{\,\prime} \|^2 \!-\! \bigl( |\eta \!-\! \eta'| \!-\! i \varepsilon \bigr)^{\!2} \Bigr] \, ,
\end{equation}
and the de Sitter breaking part~\cite{Janssen:2008px},
\begin{equation}
\mathcal{W}_{\lambda} =
	\frac{H^{D-2}}{ (4\pi)^{\frac{D}{2} } }
	\frac{ \Gamma(2\lambda) \, \Gamma(\lambda) }
		{ \Gamma\bigl( \frac{D-1}{2} \bigr) \, \Gamma\bigl( \frac{1}{2} \!+\! \lambda \bigr) }
		\frac{ e^{ ( \lambda - \frac{D-1}{2} ) u } }{ \lambda \!-\! \frac{D-1}{2} }
		\Bigl( \frac{k_0}{H} \Bigr)^{\! D-1-2\lambda }
		, \,
\label{dS breaking 2}
\end{equation}
dependent on~$u \!=\! \ln(aa')$, where $k_0<H$ is some infrared scale.

\bigskip

\noindent{\bf Subsidiary condition.}
 We derive the subsidiary condition 
for the photon propagator~\cite{Tsamis:1992zt}
by considering the expectation value of 
an anticommutator of the BRST charge operator~$\hat{\rm Q}$ 
 in Eq.~(\ref{ghost charge Q}),
with a judiciously chosen product of the vector potential and anti-ghost operators,
\begin{equation}
\bigl\{ \hat{\rm Q} , \hat{\overline{c}}(x) \hat{A}_\nu(x') \bigr\}
	=  i \Bigl[
	\nabla^\mu \! \hat{A}_\mu(x)\hat{A}_\nu(x')
	+ \xi \, \hat{\overline{c}}(x)\partial'_\nu \hat{c}(x')
	\Bigr]
	\, .
\label{BRST charge commutator}
\end{equation}
Since the BRST charge operator 
annihilates physical states,~$\hat{\rm Q} \bigl| \Omega \bigr\rangle \!=\! 0$, 
the expectation value of this anticommutator must vanish.
This produces the desired subsidiary condition,
\begin{equation}
\nabla^\mu \, i \bigl[ \tensor*[_\mu]{\Delta}{_\nu} \bigr](x;x')
	=
	-\xi \partial'_\nu \, i \Delta(x;x')
	\, ,
\label{subsidiary}
\end{equation}
relating the photon propagator to the Faddeev-Popov ghost propagator.
The photon propagator must therefore satisfy both the equation of motion~(\ref{EOMs}) 
{\it and} the subsidiary condition~(\ref{subsidiary}). The latter has seemingly gone unnoticed thus 
far, apart from the exact transverse limit~$\xi\!\to\!0$~\cite{Tsamis:2006gj}. 
This subsidiary condition is the Ward-Takahashi identity of the free theory. 
It is the massless limit of the previously derived identity for massive vector fields~\cite{Frob:2017gez} 
adapted to the Feynman 
propagator and Wightman functions.
The crucial observation is that this condition
does not admit a de Sitter invariant solution for the photon propagator,
and the reason behind it is the MMCS propagator appearing on the left hand side.
After the derivative is acted on it, 
\begin{align}
\nabla^\mu i \bigl[ \tensor*[_\mu]{\Delta}{_\nu} \bigr](x;x')
	={}&
	- \xi \bigl( \partial'_\nu y \bigr) \frac{ \partial \mathcal{F}_{\nu+1} }{ \partial y }
\nonumber \\
&
	- 
	\xi \bigl( \partial'_\nu u \bigr) 
	\frac{ H^{D-1} \, \Gamma(D \!-\! 1) }{ (4\pi)^{\frac{D}{2} } \, \Gamma\bigl( \frac{D}{2} \bigr) }
,
\label{Slavnov Taylor for our ghost propagator}
\end{align} 
the right hand side of the subsidiary condition still breaks de Sitter symmetry.
It preserves homogeneity, isotropy,
and dilations, but it breaks spatial special conformal transformations.

\bigskip

\noindent{\bf Solving for the propagator.}
Consider first the equation of motion~(\ref{EOMs}), which simplifies upon plugging 
in the Ward-Takahashi identity~(\ref{subsidiary}) for the middle term,
\begin{align}
&
\Bigl[ \square - (D\!-\!1)H^2 \Bigr] i \bigl[ \tensor*[_\mu]{\Delta}{_\nu} \bigr](x;x')
\nonumber \\
&	\hspace{1cm}
	=
	g_{\mu\nu} \frac{ i \delta^D(x \!-\!x') }{\sqrt{-g}}
	+ (1\!-\!\xi) \partial_\mu \partial'_\nu i \Delta(x;x')
	\, .
\label{simplified EOM}
\end{align}
Upon decomposing the propagator,~\footnote{
The Ward-Takahashi identity~(\ref{subsidiary}) forbids contributions 
in~(\ref{propagator decomposition: T + L}) 
that are transverse on one leg and longitudinal on the other.
}
\begin{equation}
i \bigl[ \tensor*[_\mu]{\Delta}{_\nu} \bigr](x;x')
	=
	i \bigl[ \tensor*[_\mu]{\Delta}{^{\scr \!T}_\nu} \bigr](x;x')
	+
	i \bigl[ \tensor*[_\mu]{\Delta}{^{\scr \! L}_\nu} \bigr](x;x')
	\, ,
\label{propagator decomposition: T + L}
\end{equation}
into a transverse part,
\begin{equation}
\nabla^\mu \, i \bigl[ \tensor*[_\mu]{\Delta}{^{\scr \!T}_\nu} \bigr](x;x')
	= \nabla'^\nu \, i \bigl[ \tensor*[_\mu]{\Delta}{^{\scr \!T}_\nu} \bigr](x;x') = 0
	\, ,
\end{equation}
and a longitudinal part,
\begin{equation}
i \bigl[ \tensor*[_\mu]{\Delta}{^{\scr \!L}_\nu} \bigr](x;x')
	=
	\partial_\mu \partial'_\nu L(x;x')
	\, ,
\end{equation}
the equation of motion~(\ref{EOMs})
breaks up into the transverse,
\begin{align}
&
\Bigl[ \square - (D\!-\!1)H^2 \Bigr] i \bigl[ \tensor*[_\mu]{\Delta}{^{\scr \! T}_\nu} \bigr](x;x')
\nonumber \\
&	\hspace{1.5cm}
	=
	g_{\mu\nu} \frac{ i \delta^D(x \!-\!x') }{\sqrt{-g}}
	+ \partial_\mu \partial'_\nu i \Delta(x;x')
	\, ,
\label{simplified EOM}
\end{align}
and the longitudinal equation,
\begin{equation}
\partial_\mu \partial'_\nu \square L(x;x')
	= - \xi \partial_\mu \partial'_\nu i \Delta(x;x')
	\, .
\label{longitudinal equation}
\end{equation}
The Ward-Takahashi identity~(\ref{subsidiary}) constrains only the longitudinal part,
\begin{equation}
\partial'_\nu \square L(x;x') = - \xi \, \partial'_\nu i \Delta(x;x') \, .
\label{longitudinal equation: ST}
\end{equation}
The transverse equation~(\ref{simplified EOM}) is (the massless limit of) the equation solved by 
Tsamis and Woodard~\cite{Tsamis:2006gj},
and corresponds to the propagator in the~$\xi\!\to\!0$ limit. Even though
the MMCS propagator
in the source on the right-hand side of~(\ref{simplified EOM}) 
breaks de Sitter symmetry, the two derivatives acting on it
annihilate the de Sitter breaking part.

 The longitudinal equation is, interestingly, a derivative of the subsidiary condition~(\ref{subsidiary}).
Thus, any solution satisfying the subsidiary condition will automatically satisfy the longitudinal equation of 
motion~(\ref{longitudinal equation: ST}).
But, importantly, not all solutions of the equation of motion will satisfy the subsidiary condition!
 It is the second derivative in the longitudinal equation that
enables a de Sitter-invariant solution for the longitudinal part. However,
as can be seen from~(\ref{Slavnov Taylor for our ghost propagator}), the Ward-Takahashi identity~(\ref{subsidiary})
necessitates breaking of de Sitter symmetry.

One solves Eq.~(\ref{longitudinal equation: ST}) by requiring,
\begin{equation}
\square L(x;x') = - \xi \, i \Delta(x;x') \, ,
\end{equation}
which is the equation for the so-called integrated propagator, that is solved by~\cite{Miao:2011fc},
\begin{equation}
L(x;x') =
	\frac{\xi}{2 \lambda } \frac{\partial}{\partial \lambda} i\Delta_{\lambda}(x;x')
	\bigg|_{\lambda \to \nu +1}
	\, .
\end{equation}
Combining the transverse part worked out in~\cite{Tsamis:2006gj} with
the longitudinal
part worked out here, the propagator in the general covariant gauge can be written
in a convenient covariant basis that emphasizes our main point,
\begin{align}
i \bigl[ \tensor*[_\mu]{\Delta}{_\nu} \bigr](x;x')
	={}&
	\bigl( \partial_\mu \partial'_\nu y \bigr) \, \mathcal{C}_1
	+
	\bigl( \partial_\mu y \bigr) \bigl( \partial'_\nu y \bigr) \, \mathcal{C}_2
\nonumber \\
&
	+
	\bigl( \partial_\mu u \bigr) \bigl( \partial'_\nu u \bigr) \, \mathcal{C}_4
	\, .
\label{Feynman covariant}
\end{align}
The first two terms are composed out of de Sitter invariant tensor structures
multiplied by de Sitter invariant scalar structure functions,
which depend on~$y$ only,
\begin{align}
\mathcal{C}_1 ={}&
	\frac{-1}{2\nu H^2} \biggl[
		\Bigl( \nu \!+\! \frac{1}{2} \Bigr) \mathcal{F}_\nu
		+ \Bigl( 1 \!-\! \frac{\xi}{\xi_s} \Bigr)
			\frac{\partial}{\partial y} \frac{\partial}{\partial \nu}
				\mathcal{F}_{\nu+1}
		\biggr]
		\, ,
\label{C1}
\\
\mathcal{C}_2 ={}&
	\frac{-1}{2\nu H^2} \biggl[ \frac12
		\frac{\partial }{\partial y} \mathcal{F}_\nu
		+ \Bigl( 1 \!-\! \frac{\xi}{\xi_s} \Bigr)
			\frac{\partial^2}{\partial y^2} \frac{\partial}{\partial \nu}
				\mathcal{F}_{\nu+1}
		\biggr]
		\, ,
\label{C2}
\end{align}
where $\nu$ is defined in~(\ref{MMCS propagator}),
and~$\xi_s \!=\! (D\!-\!1)/(D\!-\!3)$ is what we refer to as the simple covariant gauge.
 The last term in~(\ref{Feynman covariant}) consists of a de Sitter breaking tensor structure
multiplied by a constant,
\begin{equation}
\mathcal{C}_4 =
	\xi \times \frac{ H^{D-4} }{ (4\pi)^{\frac{D}{2}} }
	\frac{ \Gamma(D\!-\!1) }{ (D\!-\!1) \, \Gamma\bigl( \frac{D}{2} \bigr) } \, ,
\label{C4}
\end{equation}
implying that our propagator preserves dilations, but breaks special spatial conformal transformations.

All the de Sitter invariant photon propagator results reported in 
the literature~\cite{Allen:1985wd,Tsamis:2006gj,Youssef:2010dw,Frob:2013qsa} 
are captured by the~$\mathcal{C}_1$ and~$\mathcal{C}_2$ parts of the
solution.
The nonvanishing constant~$\mathcal{C}_4$ in~(\ref{C4})
however, has been overlooked thus far.
But it is this part that guarantees that the Ward-Takahashi identity
and the equation of motion are simultaneously satisfied, now producing the correct expression,
\begin{equation}
\nabla^\mu \nabla'^\nu \, i \bigl[ \tensor*[_\mu]{\Delta}{_\nu} \bigr](x;x')
	=
	- \xi \frac{ i \delta^D(x \!-\! x') }{ \sqrt{-g} }
	\,,
\label{double divergence: correct}
\end{equation}
instead of~(\ref{problem}).
Thus, expressions~(\ref{Feynman covariant}--\ref{C4})
constitute a complete solution for the Feynman propagator for the massless 
vector field, that satisfies both the equation of motion~(\ref{EOMs})
and the subsidiary condition~(\ref{subsidiary}).

The positive frequency Wightman function is now easily inferred from the solution
for the Feynman propagator by simply changing the~$i\varepsilon$
prescription for~$y$ in~(\ref{Feynman covariant})--(\ref{C2}) to the appropriate 
one,~$y_{\scr -+} \!=\! aa' \bigl[ \| \vec{x} \!-\! \vec{x}^{\,\prime} \|^2 \!-\! (\eta \!-\! \eta' \!-\! i \varepsilon)^2 \bigr]$.
This ensures that the Wightman function satisfies a homogeneous equation of motion,
\begin{equation}
\sqrt{-g} \, \mathcal{D}^{\mu\nu}
	\, i \bigl[ \tensor*[_\nu^{\scr - \!}]{\Delta}{_\alpha^{\scr \! +}} \bigr](x;x')
	=
	0
	\, ,
\label{Wightman EOM}
\end{equation}
and an appropriate subsidiary condition,
\begin{equation}
\nabla^\mu \, i \bigl[ \tensor*[^{\scr - \!}_\mu]{\Delta}{_\nu^{\scr \!+}} \bigr](x;x')
	=
	- \xi \partial'_\nu \, i \bigl[ \tensor*[^{\scr \!-\! }]{\Delta}{^{\scr \!+\! }} \bigr](x;x')
	\, ,
\end{equation}
containing the positive-frequency Wightman function for the ghost,
obtained from~(\ref{MMCS propagator}) and~(\ref{dS breaking 1})
by the same substitution~$y \!\to\! y_{\scr -+}$.
Thus, the double divergence vanishes,
\begin{equation}
\nabla^\mu \nabla'^\nu \, i \bigl[ \tensor*[_\mu^{\scr - \!}]{\Delta}{_\nu^{\scr \!+}} \bigr](x;x')
	=
	0
	\, ,
\label{double divergence: Wightman}
\end{equation}
 resolving the problems reported in~\cite{Glavan2022}.

\medskip

This result agrees with the independent
mode sum analysis in our companion
paper~\cite{Glavan2022_2}.
Apart from correctly accounting for the subsidiary condition,
the missing de Sitter breaking term resolves some issues
that we outline in the remainder of the letter.

\bigskip

\noindent{\bf Infrared behavior.}
The missing term in the photon two-point function we report here does not only solve the issue 
of non-vanishing double divergence, 
which has gone unnoticed for a long time, 
but also addresses the 
concern regarding the infrared behavior of the photon two-point function. It was reported by 
Youssef~\cite{Youssef:2010dw} (in~$D\!=\!4$) and by Rendell~\cite{Rendell:2018qid}
(in $D$ dimensions) that the two-point function does not vanish in the
deep infrared, 
\begin{equation}
i \bigl[ \tensor*[_\mu]{\Delta}{_\nu} \bigr] (x;x')
	\, \underset{!}{\overset{|y| \to  \infty}{\longsim}}
	- \xi \, 
		\frac{ H^{D-2} aa' \delta_\mu^0 \delta_\nu^0  \, \Gamma(D\!-\!1) }
			{ (4\pi)^{\frac{D}{2}} (D\!-\!1) \, \Gamma\bigl( \frac{D}{2} \bigr) }
	\, .
\end{equation}
Even though this is a gauge dependent statement~\cite{Youssef:2010dw},
since~$\xi$ ranges on the entire real line, it nonetheless is not innocuous, as it is a consequence
of failing to account for the Ward-Takahashi identity~(\ref{subsidiary}). 
This behavior is precisely removed by the missing term~(\ref{C4}) that we report,
so that in the deep infrared the photon two-point function vanishes,
\begin{equation}
i \bigl[ \tensor*[_\mu]{\Delta}{_\nu} \bigr] (x;x')
	\xrightarrow{|y| \to  \infty} 0
	\, .
\end{equation}
While this is immaterial at the linear level, it does influence the loops.

\bigskip

\noindent{\bf Energy-momentum tensor.}
Not accounting for the de Sitter breaking part~(\ref{C4}) of the photon propagator
can lead to inconsistencies in how photons source gravity.
There are two definitions possible for the photon energy-momentum tensor,
that have to coincide on-shell.
We can either define it as a variation of the gauge-invariant action,
\begin{equation}
T_{\mu\nu} = \frac{-2}{\sqrt{-g}} \frac{\delta S}{\delta g^{\mu\nu}}
	=
	\bigl( \delta_\mu^\rho \delta_\nu^\sigma \!-\! \tfrac{1}{4} g_{\mu\nu} g^{\rho\sigma} \bigr)
		g^{\alpha \beta} F_{\rho\alpha} F_{\sigma\beta}
		\, ,
\label{energy momentum: gi}
\end{equation}
or as a variation of the gauge-fixed action,
\begin{align}
T_{\mu\nu}^\star 
	={}& \frac{-2}{\sqrt{-g}} \frac{\delta S_\star}{\delta g^{\mu\nu}}
	= T_{\mu\nu} - \frac{2}{\xi} A_{(\mu} \nabla_{\nu)} \nabla^\rho \! A_\rho
\nonumber \\
&
	+ \frac{ g_{\mu\nu} }{ \xi } \Bigl[ A_\rho \nabla^\rho \nabla^\sigma \! A_\sigma
		+ \tfrac{1}{2} \bigl( \nabla^\rho \! A_\rho \bigr)^{\!2\,} \Bigr] 
\nonumber \\
& 
	- 2  \bigl( \partial_{(\mu}\overline{c} \bigr) \bigl( \partial_{\nu)} c \bigr)
	+ g_{\mu\nu}  g^{\rho\sigma} \bigl( \partial_\rho \overline{c} \bigr) \bigl( \partial_\sigma c \bigr)
\, .
\label{energy momentum: *}
\end{align}
Classically the two give the same answer as they differ by BRST-exact terms only, which all vanish 
on-shell~\footnote{
BRST-exact objects are the ones that can be written as (anti)commutators
of other objects with the BRST charge. They vanish on-shell by the definition of the formalism.
}.
The quantized theory has to maintain this property at the level of expectation values.
Operators associated to the two definitions of the energy-momentum tensor
are defined by Weyl ordered (symmetrized) products of field operators,
and the difference between their expectation values is best
expressed in terms of the Wightman function and derivatives acting on it,
\begin{align}
\MoveEqLeft[2]
\bigl\langle \hat{T}_{\mu\nu}^\star(x) \bigr\rangle 
	- \bigl\langle \hat{T}_{\mu\nu}(x) \bigr\rangle\nonumber 
	=
	\biggl\{ 
	\frac{ g_{\mu\nu} }{ 2\xi } \nabla^\rho \nabla'^\sigma
	i \bigl[ \tensor*[_\rho^{\scr - \! }]{\Delta}{_\sigma^{\scr \! +} } \bigr] (x;x')
\nonumber \\
&
	- \frac{1}{\xi} \Bigl[ 
		\delta^\rho_{(\mu} \nabla'_{\nu)} \nabla'^\sigma
		+ \delta^\sigma_{(\mu} \nabla_{\nu)} \nabla^\rho \Bigr]
	  i \bigl[ \tensor*[_\rho^{\scr - \!}]{\Delta}{_\sigma^{\scr \! +}} \bigr] (x;x')
\nonumber \\
&	
	+ \frac{g_{\mu\nu} }{ 2\xi }
	\Bigl[
	\nabla'^\rho \nabla'^\sigma 
	+ \nabla^\sigma \nabla^\rho 
	\Bigr] 
	 i \bigl[ \tensor*[_\rho^{\scr - \!}]{\Delta}{_\sigma^{\scr \! +}} \bigr] (x;x')
\nonumber \\
&	
	- \Bigl[ 2 \nabla_\mu \nabla'_\nu - g_{\mu\nu} \nabla^\rho \nabla'_\rho \Bigr] 
	i \bigl[ \tensor*[^{\scr \!-\! }]{\Delta}{^{\scr \!+\! }} \bigr](x;x')
	\biggr\} {\bigg|}_{x' \to x}
	\, .
\label{T exp}
\end{align}
The terms in the last three lines above cancel on the account of
 a derivative of the Ward-Takahashi identity,
\begin{equation}
\nabla_\rho \nabla^\mu i \bigl[ \tensor*[_\mu^{\scr - \!}]{\Delta}{_\nu^{\scr \!+}} \bigr](x;x')
	= - \xi \partial_\rho \partial'_\nu \,
		i \bigl[ \tensor*[^{\scr \!-\! }]{\Delta}{^{\scr \!+\! }} \bigr](x;x')
	\, ,
\end{equation}
which is insensitive to the de Sitter breaking part that drops out because of the second derivative.
The remaining term from the first line,
\begin{equation}
\bigl\langle \hat{T}_{\mu\nu}^\star(x) \bigr\rangle 
	- \bigl\langle \hat{T}_{\mu\nu}(x) \bigr\rangle
\\
	=
	\frac{g_{\mu\nu} }{ 2\xi }
	\nabla^\rho \nabla'^\sigma
	 i \bigl[ \tensor*[_\rho^{\scr - \!}]{\Delta}{_\sigma^{\scr \! +}} \bigr] (x;x')
	\Big|_{x' \to x}
	\, ,
\label{Tmn: extra term}
\end{equation}
has to vanish on its own.
This is precisely in the form of the problematic expression~(\ref{problem}) we started with.
If we were to disregard the de Sitter breaking part~(\ref{C4}) of the photon propagator,
this difference would not vanish,
\begin{equation}
\bigl\langle \hat{T}_{\mu\nu}^\star(x) \bigr\rangle 
	- \bigl\langle \hat{T}_{\mu\nu}(x) \bigr\rangle\nonumber \\
	\overset{!}{=}
	- \frac{g_{\mu\nu} }{ 2 }
	\frac{ H^{D} \, \Gamma(D) }{ (4\pi)^{\frac{D}{2} } \, \Gamma\bigl( \frac{D}{2} \bigr) } 
	\, .
\end{equation}
but would have a cosmological constant form.
Since this contribution is independent of the gauge-fixing parameter~$\xi$, it 
 would be difficult
to recognize it as an unphysical answer, 
even though it clearly originates from 
 not accounting for the gauge sector constraints.
In fact, when one uses the de Sitter breaking two-point function, 
inserting~(\ref{double divergence: Wightman}) into~(\ref{Tmn: extra term}) 
 shows that the two definitions give the same answer,
\begin{equation}
\bigl\langle \hat{T}_{\mu\nu}^\star(x) \bigr\rangle 
	- \bigl\langle \hat{T}_{\mu\nu}(x) \bigr\rangle
	=
	0
	\, .
\end{equation}
When computed in dimensional regularization the expectation value of the energy-momentum tensor
in~$D\!=\!4$ vanishes,~$\langle \hat{T}_{\mu\nu}(x) \rangle \!=\! 0$, as shown in~\cite{Glavan2022_2}.

\bigskip


\noindent{\bf Discussion.}
The photon propagator in de Sitter in the general covariant gauge
takes the form~(\ref{Feynman covariant}) with the three structure functions given in~(\ref{C1}--\ref{C4}).
The de Sitter invariant parts in~(\ref{C1}--\ref{C2})
have been derived in previous works~\cite{Allen:1985wd,Tsamis:2006gj,Youssef:2010dw,Frob:2013qsa}.
It is the nonvanishing constant~$\mathcal{C}_4$ in~(\ref{C4}), overlooked thus
far, that is our main contribution. This contribution {\it breaks} de Sitter symmetry. In particular,
it breaks special spatial conformal transformations, while preserving dilations, 
spatial homogeneity and isotropy. Even though this term is a homogeneous solution 
of the propagator equation of motion~(\ref{EOMs}),  it cannot be 
discarded. It is the Ward-Takahashi identity~(\ref{subsidiary}), necessary for 
the construction of
a consistent propagator,
that requires it. The de Sitter breaking can be traced back to the Faddeev-Popov ghost propagator,
which satisfies the equation for the massless, minimally coupled scalar~(\ref{ghost EOMs}), 
which does not admit a de Sitter invariant solution. This ghost propagator appears in the Ward-Takahashi identity~(\ref{subsidiary}), whose solution allows a dilation-preserving photon propagator. At the linear level,
this necessary modification of the photon propagator is of little physical significance,
as it is confined to the pure gauge sector. However, it can be
of paramount importance
when interactions are considered, as failure to implement it, 
in general leads to incorrect results.

\bigskip
\noindent {\bf Acknowledgements.} 
D.G. was supported by the Czech Science 
Foundation (GA\v{C}R) grant No.~20-28525S.
This work is part of the Delta ITP consortium, a program of the Netherlands Organisation
for Scientific Research (NWO) that is funded by the Dutch Ministry of Education, Culture
and Science (OCW) -- NWO project number 24.001.027.

\end{document}